\documentclass[journal]{IEEEtran}
\usepackage{amssymb}
\usepackage{amsmath}
\usepackage{graphicx}
\usepackage{booktabs}
\usepackage[hidelinks]{hyperref}
\usepackage{multirow}
\usepackage{pifont}
\usepackage{float}
\usepackage[section]{placeins}
\usepackage[ruled,vlined,linesnumbered]{algorithm2e}
\usepackage{tabularx}
\usepackage{array}
\usepackage{threeparttable}
\usepackage{amsthm}
\usepackage{algorithmic}
\usepackage{enumitem}
\usepackage{adjustbox}
\usepackage{makecell}
\newtheorem{definition}{Definition}
\providecommand{\etal}{\textit{et al.}}
\providecommand{\eg}{e.g.}

\begin{document}

\title{Controllability-Aware Adversarial Examples Against LLM-Based Network Traffic Classifiers}

\author{Zhenpeng~Li%
\thanks{Z.~Li is with Guangzhou Health Science College,
GuangYuanZhong Road~248, Guangzhou, Guangdong~510405, China
(e-mail: 2025301001@gzws.edu.cn). Corresponding author.
Manuscript received ---; revised ---; accepted ---.}}

\maketitle

\begin{abstract}
Large language models (LLMs) are increasingly explored as network intrusion detection classifiers, but their adversarial robustness under realistic attacker constraints remains unclear. We present a controllability-aware black-box transfer framework for LLM-based network traffic classifiers. The framework partitions flow features into directly controllable (DC), indirectly controllable (IC), and uncontrollable (UC) groups according to network communication semantics, then restricts perturbations to DC features while freezing IC/UC features. Using a shared XGBoost surrogate, we generate finite-difference PGD, greedy coordinate-wise, and NES adversarial examples and transfer them to seven LLM targets and two conventional ML targets across five IDS benchmarks from 1999 to 2022. Across 27 valid LLM configurations and over 500,000 adversarial examples, we find that LLM transfer vulnerability is substantial but dataset- and comparator-dependent. Compared with LightGBM, LLMs are more vulnerable on RT-IoT2022 and CIC-IDS-2018, comparable on NSL-KDD and UNSW-NB15, and less vulnerable on HIKARI-2021; compared with the averaged ML baseline, LLMs show higher ASR on all five datasets. We further observe a consistent cross-architecture transfer hierarchy: gradient- and score-based perturbations transfer more effectively than greedy perturbations across all 27 LLM cells and 9/10 ML cells. Cross-surrogate validation with tree, neural, and linear surrogates yields similar LLM ASR, reducing evidence that the findings are XGBoost-specific. Constraint violation rate is 0\% by construction.
\end{abstract}


\IEEEpeerreviewmaketitle

\section{Introduction}
\label{sec:intro}

The use of large language models for network intrusion detection is an active research direction, driven by their ability to process heterogeneous flow features as natural language descriptions~\cite{ferrag2024revolutionizing, xu2024large}.
While production deployment remains limited, commercial systems such as Microsoft Copilot for Security already integrate LLMs into security operations pipelines, processing network telemetry alongside threat intelligence.
Whether LLM-based classifiers can be evaded through controllability-constrained traffic modifications is therefore a concrete security question: production-integrated LLMs processing live network telemetry can be targeted by adversaries with access only to public data and a local classifier.

Adversarial robustness of traditional ML-based intrusion detection systems (IDS) has been studied extensively~\cite{apruzzese2022modeling, han2021evaluating, pierazzi2020intriguing}.
However, existing work suffers from two limitations when applied to the LLM-IDS setting.
First, most attacks perturb arbitrary features without respecting the attacker's operational constraints---an attacker cannot control response bytes from a remote server, nor dictate aggregated statistics computed by network monitors.
Second, no prior work evaluates whether adversarial transferability---the dominant practical attack vector---behaves differently when the target is a Transformer-based language model versus a gradient-boosted tree or neural network.

We address both gaps through three contributions:

\textbf{(C1) Controllability-Aware Attack Framework.}
We formalize an attacker capability taxonomy that classifies network flow features into Directly Controllable~(DC), Indirectly Controllable~(IC), and Uncontrollable~(UC) categories, grounded in the physical semantics of network communication (Section~\ref{sec:threat}).
Adversarial perturbations are restricted to DC features, producing examples that are more realistic than unconstrained feature-space attacks, though we note that full protocol-level validity requires additional constraints beyond the scope of this work (Section~\ref{sec:constraint_limits}).

\textbf{(C2) Cross-Architecture Robustness Evaluation.}
We conduct a systematic comparison of LLM and traditional ML classifier robustness under a unified black-box transfer threat model.
Our evaluation spans five datasets (NSL-KDD through RT-IoT2022), seven LLMs, and two ML classifiers, producing 27 valid LLM evaluation cells across five random seeds (Section~\ref{sec:results}).
We report results disaggregated by ML target type, perturbation budget, and classifier competence level to control for potential confounds.

\textbf{(C3) Cross-Architecture Transfer Hierarchy.}
We identify and empirically validate a consistent pattern: gradient-based~(PGD) and score-based~(NES) perturbation algorithms generate adversarial examples with higher cross-architecture transferability than greedy coordinate-wise perturbations (Section~\ref{sec:analysis}).
This pattern holds across all 27 LLM cells; an 8-cell cross-surrogate check spanning all five datasets reduces the likelihood that it is specific to the XGBoost surrogate.
We quantify the underlying mechanism using prediction agreement as a proxy for architectural similarity.

In the evaluated transfer setting, LLM vulnerability is substantial but dataset-dependent.
Against the averaged ML baseline (LightGBM + DNN-IDS), LLMs show higher ASR on all five datasets (dataset-level Wilcoxon $p=0.031$, Cohen's $d=1.21$).
Disaggregated against LightGBM specifically, LLMs are more vulnerable on CIC-IDS-2018 ($1.4\times$) and RT-IoT2022 ($7.2\times$), comparable on NSL-KDD and UNSW-NB15, and less vulnerable on HIKARI-2021 (LightGBM $1.43\times$ more vulnerable, DC\,=\,25).
The gap on modern high-traffic datasets persists when restricted to high-competence classifiers (clean F1~$\geq 0.70$); this high-competence subset is the primary basis for vulnerability claims, with the full F1~$\geq 0.45$ set reported for completeness.

\section{Related Work}
\label{sec:related}

\subsection{Controllability in IDS Adversarial Evaluation}

Adversarial IDS research has established that unconstrained feature-space perturbations can overstate practical evasion risk~\cite{goodfellow2015explaining, madry2018towards}.
Pierazzi \etal~\cite{pierazzi2020intriguing} make this point through the distinction between feature-space and problem-space attacks: an adversarial feature vector is meaningful only if it can correspond to realizable traffic.
Apruzzese \etal~\cite{apruzzese2022modeling} similarly emphasize attacker capability constraints in network security evaluations.
These studies motivate a stricter evaluation setting, but they do not by themselves specify which flow features a source-side attacker can directly change, which features can only be influenced indirectly, and which features are outside the attacker's control.

Traffic-manipulation studies move closer to operational evasion.
Han \etal~\cite{han2021evaluating} demonstrate that ML-based traffic classifiers can be evaded through practical traffic changes, while Hashemi and Keller~\cite{hashemi2020enhancing} and Peng \etal~\cite{peng2019adversarial} study robustness for specific IDS architectures.
For the question addressed in this paper, however, two limitations remain.
First, the attack surface is usually not expressed as a feature-level controllability taxonomy.
Second, the target is typically a conventional ML classifier rather than a text-serialized LLM classifier evaluated under the same transfer attack protocol.

\subsection{LLM-Based Intrusion Detection}

Recent work has explored using LLMs as network traffic classifiers by serializing flow features into natural language prompts~\cite{ferrag2024revolutionizing,xu2024large}.
More broadly, tabular-to-text prompting has been studied for few-shot tabular classification~\cite{hegselmann2022tabllm}, supporting the representation shift from numeric feature vectors to token sequences.
This line of work changes the target representation: tabular flow records are converted into token sequences, and classification is performed by a language model rather than by a model trained directly on numeric feature vectors.
That change creates a robustness question that accuracy-oriented evaluations do not answer.
If attacks are generated on a local tabular surrogate and transferred to a deployed target, the relevant comparison is not simply ``LLM versus ML'' in aggregate.
The comparison depends on whether the ML reference is a tree ensemble, a neural classifier, or an average over architectures with different transfer behavior.

\subsection{Black-Box Transfer and Target-Architecture Effects}

Transfer-based attacks~\cite{papernot2017practical, liu2017delving} craft adversarial examples on a surrogate model and apply them to a target without direct access.
Score-based methods such as NES~\cite{ilyas2018black} estimate gradients through finite differences on model outputs.
Greedy coordinate-wise attacks~\cite{chen2019hopskipjumpattack} perturb features sequentially based on local improvement.
These methods are usually discussed as different attack families, but in a black-box IDS deployment they also provide a way to test whether perturbations transfer similarly across target architectures.
This distinction is central to the present study: a tree-to-tree transfer path, a tree-to-neural transfer path, and a tree-to-LLM transfer path can produce different conclusions about vulnerability.
An averaged ML baseline can therefore obscure the comparator-dependent pattern that matters for assessing LLM-based IDS.

\subsection{Comparison with Prior Work}
\label{sec:related_comparison}

Table~\ref{tab:related_comparison} positions the present work against the closest prior studies along six dimensions critical for a realistic, reproducible evaluation of adversarial robustness in network IDS.

\begin{table}[!t]
\centering
\caption{Comparison with closest prior work across six evaluation dimensions.
\checkmark\,=\,fully addressed; $\circ$\,=\,partial or informal; ---\,=\,absent.}
\label{tab:related_comparison}
\footnotesize
\setlength{\tabcolsep}{2.2pt}
\renewcommand{\arraystretch}{1.08}
\begin{tabular}{@{}p{0.31\columnwidth}cccccc@{}}
\toprule
\textbf{Study} 
& \makecell{\textbf{Ctrl.}\\\textbf{Tax.}} 
& \textbf{LLM} 
& \makecell{\textbf{LLM}\\\textbf{vs.\ ML}} 
& \makecell{\textbf{BB}\\\textbf{Trans.}} 
& \makecell{\textbf{Multi-}\\\textbf{DS}} 
& \makecell{\textbf{PS}\\\textbf{Valid.}} \\
\midrule
Pierazzi~\etal~\cite{pierazzi2020intriguing}  
& $\circ$ & --- & --- & --- & --- & \checkmark \\
Apruzzese~\etal~\cite{apruzzese2022modeling}  
& $\circ$ & --- & --- & --- & --- & $\circ$ \\
Han~\etal~\cite{han2021evaluating}            
& --- & --- & --- & \checkmark & $\circ$ & \checkmark \\
Ferrag~\etal~\cite{ferrag2024revolutionizing} 
& --- & \checkmark & --- & --- & \checkmark & --- \\
Xu~\etal~\cite{xu2024large}                   
& --- & \checkmark & --- & --- & $\circ$ & --- \\
\textbf{This work}                            
& \checkmark & \checkmark & \checkmark & \checkmark & \checkmark & $\circ$ \\
\bottomrule
\end{tabular}

\vspace{2pt}
\begin{minipage}{\columnwidth}
\scriptsize
\textit{Columns}: Ctrl.\ Tax.\ = formal attacker controllability taxonomy; 
LLM = LLM-IDS as attack target; 
LLM\,vs.\,ML = cross-architecture vulnerability comparison; 
BB Trans.\ = black-box transfer threat model; 
Multi-DS = five benchmark datasets; 
PS Valid.\ = problem-space validity. 
Our work enforces feature-group controllability and verifies CVR\,=\,0\%, 
but does not enforce full traffic replay validity.
\end{minipage}
\end{table}

The table shows that prior work addresses important parts of the problem, but not the combination needed to evaluate the central claim of this paper.
Problem-space and capability-aware IDS studies establish why unconstrained feature attacks are insufficient, but they do not evaluate LLM-based targets or disaggregate LLM-vs-ML transfer by target architecture.
LLM-based IDS studies establish the relevance of language-model classifiers, but their evaluations focus on detection rather than controllability-constrained adversarial transfer.
The gap is therefore not the absence of adversarial IDS research or the absence of LLM-IDS research.
The gap is the missing intersection: feature-level attacker controllability, black-box transfer, LLM targets, ML comparators, and multi-dataset disaggregated comparison in a single evaluation.
Our validity guarantee remains narrower than full problem-space replay: the attacks respect feature controllability, while type constraints, inter-feature dependencies, and attack functionality remain limitations.

\section{Threat Model and Formalization}
\label{sec:threat}

\begin{figure*}[!t]
\centering
\vspace{-4pt}
\includegraphics[width=0.82\textwidth,trim={0.05in 0.20in 0.05in 0.10in},clip]{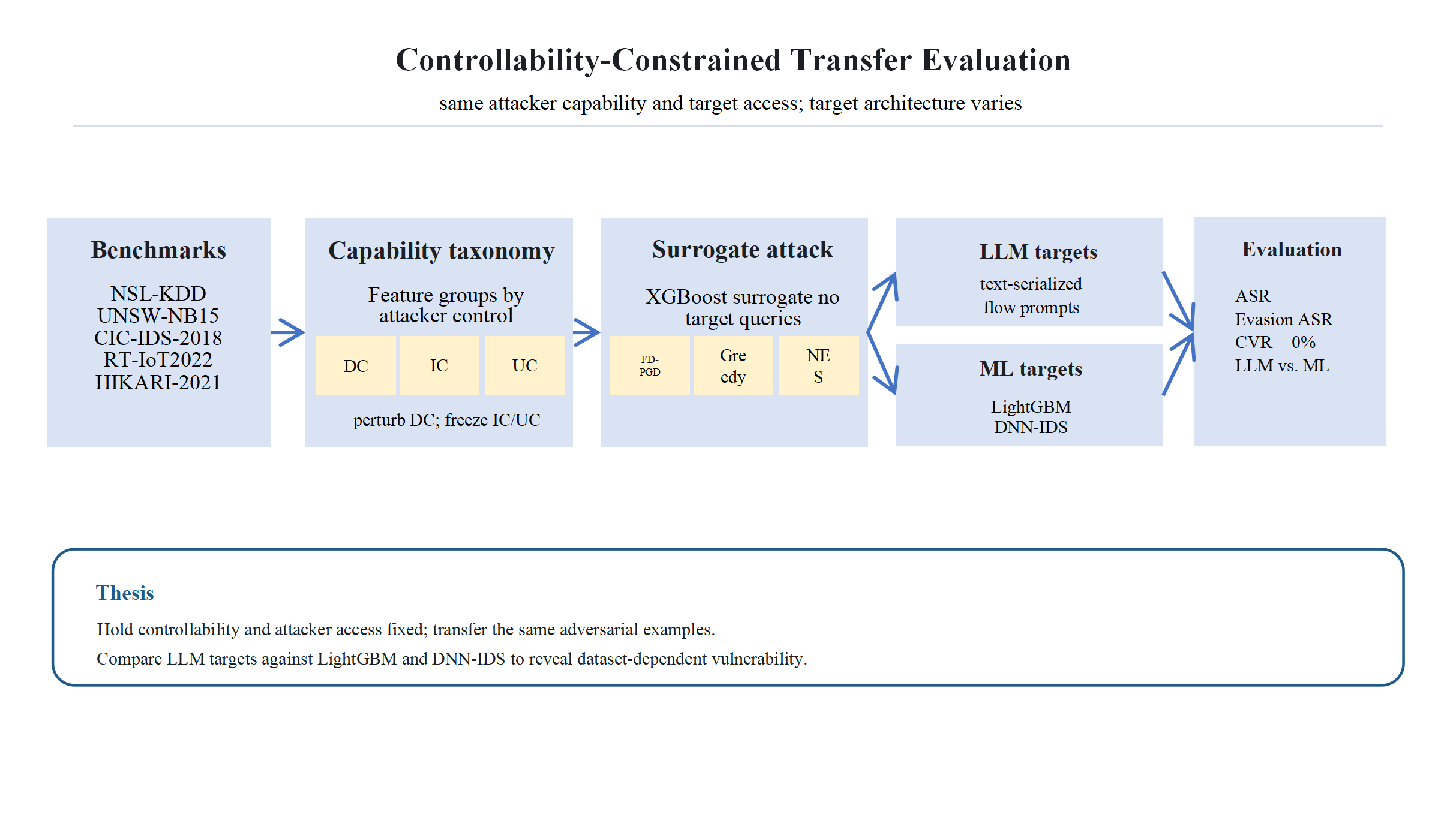}
\vspace{-4pt}
\caption{Controllability-aware black-box transfer evaluation framework. 
Flow features are partitioned into DC/IC/UC groups; only DC features are perturbed on a shared XGBoost surrogate and transferred to LLM and ML IDS targets under identical attacker constraints.}
\label{fig:framework}
\vspace{-6pt}
\end{figure*}

The framework fixes attacker capability (DC/IC/UC taxonomy), surrogate access (black-box transfer), and perturbation structure (FD-PGD, greedy, NES), ensuring that LLM-vs-ML ASR differences are attributable to target architecture rather than to varying attack conditions.
Fig.~\ref{fig:framework} summarizes the pipeline.

\subsection{Attacker Capability Taxonomy}

The first requirement is to define the attacker's controllable feature subspace before generating adversarial examples.
We model an attacker who controls the source endpoint of a network flow and seeks to evade detection while preserving attack functionality.
This attacker can directly set some source-side features, indirectly influence some aggregate statistics, and cannot control features determined by the remote endpoint, network path, or monitor.

\begin{definition}[Feature Controllability]
\label{def:controllability}
Given a network flow feature vector $\mathbf{x} \in \mathbb{R}^d$, we partition the feature index set $\{1, \ldots, d\}$ into three disjoint subsets:
\begin{itemize}
    \item \textbf{Directly Controllable (DC):} Features determined solely by the attacker's endpoint behavior (\eg, source bytes, packet count, TCP flags set by sender).
    \item \textbf{Indirectly Controllable (IC):} Aggregated statistics over multiple flows that the attacker can influence through sustained behavioral changes (\eg, connection rates, service ratios).
    \item \textbf{Uncontrollable (UC):} Features determined by the remote endpoint, network infrastructure, or monitoring system (\eg, destination bytes, response times, round-trip delays).
\end{itemize}
\end{definition}

Let $\mathcal{I}_{\text{DC}}$, $\mathcal{I}_{\text{IC}}$, $\mathcal{I}_{\text{UC}}$ denote the index sets, with $\mathcal{I}_{\text{DC}} \cup \mathcal{I}_{\text{IC}} \cup \mathcal{I}_{\text{UC}} = \{1, \ldots, d\}$ and pairwise disjoint.
Table~\ref{tab:feature_taxonomy} applies the same taxonomy to each dataset.
The resulting DC counts differ by dataset, which is important for the central thesis: transfer vulnerability is evaluated under dataset-specific controllable subspaces rather than under a uniform unconstrained feature space.

\begin{table}[!t]
\centering
\caption{Feature controllability taxonomy across datasets. $|\cdot|$ denotes count; representative features shown for each category.}
\label{tab:feature_taxonomy}
\small
\begin{tabular}{l c c c}
\toprule
\textbf{Dataset} & $|\mathcal{I}_{\text{DC}}|$ & $|\mathcal{I}_{\text{IC}}|$ & $|\mathcal{I}_{\text{UC}}|$ \\
\midrule
NSL-KDD (41) & 20 & 9 & 12 \\
\footnotesize\textit{src\_bytes, flags} & \footnotesize\textit{srv\_count} & \footnotesize\textit{dst\_bytes} \\[2pt]
UNSW-NB15 (41) & 15 & 11 & 15 \\
\footnotesize\textit{sbytes, spkts} & \footnotesize\textit{sload, rate} & \footnotesize\textit{dbytes, tcprtt} \\[2pt]
CIC-IDS-2018 (25) & 14 & 7 & 4 \\
\footnotesize\textit{Tot Fwd Pkts} & \footnotesize\textit{Flow IAT} & \footnotesize\textit{Bwd Pkt Len} \\[2pt]
RT-IoT2022 (83) & 27 & 32 & 24 \\
\footnotesize\textit{fwd\_pkts, proto} & \footnotesize\textit{flow\_iat, active} & \footnotesize\textit{bwd\_pkts, bwd\_iat} \\[2pt]
HIKARI-2021 (82) & 25 & 32 & 25 \\
\footnotesize\textit{fwd\_pkts, originp} & \footnotesize\textit{flow\_iat, active} & \footnotesize\textit{bwd\_pkts, bwd\_iat} \\
\bottomrule
\end{tabular}
\end{table}

\subsection{Semantic Constraint Space}

The taxonomy becomes operational only when it is enforced during perturbation.
The constraint space below defines the admissible adversarial examples used throughout the evaluation.

\begin{definition}[Controllability Constraint Space]
\label{def:constraint}
Given a clean sample $\mathbf{x}$ and perturbation budget $\varepsilon > 0$, the controllability constraint space is:
\begin{equation}
\label{eq:constraint}
\mathcal{S}(\mathbf{x}, \varepsilon) = \left\{ \mathbf{x}' \in [0,1]^d \;\middle|\;
\begin{array}{l}
\|\mathbf{x}'_{\mathcal{I}_{\text{DC}}} - \mathbf{x}_{\mathcal{I}_{\text{DC}}}\|_\infty \leq \varepsilon \\[3pt]
\mathbf{x}'_{\mathcal{I}_{\text{IC}}} = \mathbf{x}_{\mathcal{I}_{\text{IC}}} \\[3pt]
\mathbf{x}'_{\mathcal{I}_{\text{UC}}} = \mathbf{x}_{\mathcal{I}_{\text{UC}}}
\end{array}
\right\}
\end{equation}
where $\mathbf{x}'_{\mathcal{I}}$ denotes the subvector indexed by $\mathcal{I}$, and all features are in min-max normalized $[0,1]$ space.
\end{definition}

The set $\mathcal{S}$ allows perturbations only on DC features and freezes IC/UC features.
This makes every LLM and ML target receive adversarial examples generated under the same attacker-capability boundary.
The constraint violation rate~(CVR) is zero by construction: it checks that the DC mask does not inadvertently modify IC/UC features.
We report CVR as an implementation check, not as an independent empirical finding.

\subsubsection{Scope and Limitations of the Constraint Model}
\label{sec:constraint_limits}

Our controllability constraint enforces \emph{feature-group masking}: only DC features are perturbed within their normalized range.
This is a necessary but not sufficient condition for full network-level realizability.
Specifically, our framework does not enforce:
(i)~type constraints (integer features such as packet counts may receive non-integer perturbations in normalized space);
(ii)~inter-feature dependencies (\eg, the relationship between \texttt{src\_bytes} and \texttt{mean\_pkt\_len});
(iii)~attack-functionality preservation (whether the perturbed flow retains its malicious behavior).
Our results should therefore be interpreted as evaluating robustness to \emph{controllability-constrained feature-space perturbations}---closer to attacker capability than unconstrained attacks, but not equivalent to end-to-end traffic replay.
Extending the constraint model to enforce type and dependency validity is an important direction for future work.

\subsection{Black-Box Transfer Threat Model}

The second requirement is to keep target access constant across LLM and ML targets.
We therefore use a transfer setting in which the attacker trains a local surrogate and never queries the deployed target.

\begin{definition}[Threat Model]
\label{def:threat}
The attacker has:
\begin{enumerate}
    \item Access to a labeled training set $\mathcal{D}_{\text{train}}$ (public benchmark data).
    \item The ability to train a surrogate classifier $f_S$ on $\mathcal{D}_{\text{train}}$.
    \item No access to the target classifier $f_T$---neither architecture, parameters, nor query interface.
\end{enumerate}
\end{definition}

This formulation captures a deployment setting in which the attacker knows the extracted features and can train a local model on public or matched-distribution data, but cannot probe the deployed classifier.
Both LLM and ML targets are evaluated on identical adversarial examples generated from the same surrogate.
This is the mechanism that makes disaggregated target-architecture comparison meaningful: differences in ASR can be attributed to transfer behavior under the tested protocol rather than to different attacker access.

The primary surrogate is an XGBoost classifier trained on the same data partition.
It serves exclusively as the perturbation generation platform and is never evaluated as a target.
In Section~\ref{sec:surrogate}, we validate that conclusions hold with MLP and logistic regression surrogates.

\subsection{Perturbation Generation Algorithms}

The third requirement is to separate the target-architecture question from the perturbation-generator question.
We therefore apply three algorithms to the same surrogate $f_S$, with each algorithm exploiting different surrogate information.
All three operate within $\mathcal{S}(\mathbf{x}, \varepsilon)$; none requires access to the target~$f_T$.

\textbf{FD-PGD (gradient-based).}
Projected Gradient Descent~\cite{madry2018towards} with 20 steps, step size $\alpha = 2.5\varepsilon/n_{\text{steps}}$, restricted to DC features.
Since tree-ensemble models do not expose analytical input gradients, we estimate $\nabla_{\mathbf{x}} \mathcal{L}$ via forward finite differences with step size $\delta = 10^{-3}$:
\begin{equation}
\hat{g}_j = \frac{\mathcal{L}(f_S, \mathbf{x} + \delta\mathbf{e}_j, y) - \mathcal{L}(f_S, \mathbf{x}, y)}{\delta}, \quad j \in \mathcal{I}_{\text{DC}}
\end{equation}
The PGD update then proceeds with the standard sign-gradient step:
\begin{equation}
\mathbf{x}^{(t+1)}_{\mathcal{I}_{\text{DC}}} = \Pi_{\mathcal{S}} \left[ \mathbf{x}^{(t)}_{\mathcal{I}_{\text{DC}}} + \alpha \cdot \mathrm{sign}(\hat{\mathbf{g}}) \right]
\end{equation}
This zeroth-order PGD variant is standard for non-differentiable surrogates~\cite{chen2017zoo} and produces perturbations comparable to exact-gradient PGD on differentiable models.

\textbf{Greedy (coordinate-wise).} Iteratively selects the single DC feature yielding maximum loss increase on $f_S$, 10 steps, step size $\varepsilon/2$:
\begin{equation}
j^* = \arg\max_{j \in \mathcal{I}_{\text{DC}}} \; \mathcal{L}(f_S, \mathbf{x} + \varepsilon \cdot \mathbf{e}_j, y) - \mathcal{L}(f_S, \mathbf{x}, y)
\end{equation}

\textbf{NES (score-based).} Natural Evolution Strategies~\cite{ilyas2018black} estimates gradients of $f_S$'s output via antithetic sampling with 20 query samples per step, 10 steps, smoothing $\sigma = 0.01$, restricted to DC indices.

All three algorithms query exclusively the surrogate~$f_S$.
Their role is not to define separate threat models, but to test whether different perturbation structures survive transfer across target architectures.
FD-PGD uses per-feature finite differences and aggregates information across all DC features simultaneously; NES uses stochastic random-direction finite differences; greedy uses discrete single-coordinate evaluations.

\textbf{Query budget comparison.}
We calibrate parameters so that algorithm comparisons are not driven primarily by surrogate-query budget: FD-PGD requires $20 \times (|\mathcal{I}_{\text{DC}}| + 1)$ queries (20~steps, $|\mathcal{I}_{\text{DC}}|$~forward differences per step), greedy requires $10 \times 2|\mathcal{I}_{\text{DC}}|$ (10~steps, $\pm$~direction per DC feature), and NES requires $10 \times (2 \times 20 + 1) = 410$ queries (10~steps, 20~antithetic pairs).
For NSL-KDD ($|\mathcal{I}_{\text{DC}}| = 20$), this yields 420, 400, and 410 queries respectively---within $5\%$ of each other.
For datasets with larger DC sets (RT-IoT2022, $|\mathcal{I}_{\text{DC}}|=27$: FD-PGD\,=\,560, greedy\,=\,540, NES\,=\,410; HIKARI-2021, $|\mathcal{I}_{\text{DC}}|=25$: 520/500/410), NES uses 25--27\% fewer surrogate queries than FD-PGD and greedy, yet achieves comparable or higher ASR.
This imbalance favors NES conservatively: reported NES results are obtained with a \emph{smaller} query budget, so if anything the comparison underestimates NES efficiency relative to FD-PGD and greedy.

\subsection{Cross-Architecture Transfer Hierarchy}

The final method component states the transfer pattern that the experiments are designed to test.
If LLM vulnerability is comparator-dependent, then the evaluation must explain not only whether attacks transfer, but also when transfer changes with surrogate-target architectural similarity.

\begin{quote}
\textbf{Empirical Finding~1 (Transfer Hierarchy).}
\label{find:hierarchy}
\textit{Under black-box transfer from a surrogate $f_S$ to an architecturally dissimilar target $f_T$, gradient-based (FD-PGD) and score-based (NES) perturbation algorithms produce adversarial examples with higher transferability than greedy coordinate-wise perturbations.
The hierarchy weakens as surrogate-target similarity increases, measured by prediction agreement rate $\rho(f_S, f_T)$.}
\end{quote}

We quantify architectural similarity using the prediction agreement rate $\rho(f_S, f_T) = \frac{1}{N}\sum_{i=1}^{N} \mathbf{1}[f_S(\mathbf{x}_i) = f_T(\mathbf{x}_i)]$ on the evaluation set.
In our experiments, the hierarchy holds for all cells where $\rho < 0.95$ and reverses only in the single cell where $\rho = 0.98$ (CIC-IDS-2018, XGBoost$\to$LightGBM).
Section~\ref{sec:mechanism} provides a detailed mechanistic analysis.

\textit{Rationale.}
Greedy perturbations exploit model-specific decision boundary geometry---they find the single most effective coordinate on~$f_S$.
This specificity yields high on-surrogate effectiveness but poor cross-architecture transfer.
FD-PGD and NES aggregate information across multiple coordinates simultaneously, producing perturbation directions that align with loss-landscape features less specific to a single surrogate.

The overall attack procedure is summarized in Algorithm~\ref{alg:attack}.

\begin{algorithm}[!t]
\caption{Controllability-Constrained Black-Box Transfer Attack}
\label{alg:attack}
\begin{algorithmic}[1]
\REQUIRE Dataset $\mathcal{D}$, taxonomy $(\mathcal{I}_{\text{DC}}, \mathcal{I}_{\text{IC}}, \mathcal{I}_{\text{UC}})$, budget $\varepsilon$, algorithm $\mathcal{A} \in \{\text{FD-PGD}, \text{greedy}, \text{NES}\}$
\ENSURE Attack success rate against target $f_T$
\STATE Train surrogate $f_S$ (XGBoost) on $\mathcal{D}_{\text{train}}$
\STATE Select evaluation set $\mathcal{D}_{\text{eval}}$ (500/class, stratified)
\FOR{each $(\mathbf{x}_i, y_i) \in \mathcal{D}_{\text{eval}}$}
    \STATE $\mathbf{x}'_i \leftarrow \mathcal{A}(f_S, \mathbf{x}_i, y_i, \varepsilon, \mathcal{I}_{\text{DC}})$ \COMMENT{perturb on surrogate}
    \STATE $\mathbf{x}'_i[\mathcal{I}_{\text{IC}} \cup \mathcal{I}_{\text{UC}}] \leftarrow \mathbf{x}_i[\mathcal{I}_{\text{IC}} \cup \mathcal{I}_{\text{UC}}]$ \COMMENT{freeze}
    \STATE $\mathbf{x}'_i \leftarrow \mathrm{clip}(\mathbf{x}'_i, 0, 1)$
\ENDFOR
\STATE $\text{ASR} \leftarrow \frac{|\{i : f_T(\mathbf{x}_i) = y_i \wedge f_T(\mathbf{x}'_i) \neq y_i\}|}{|\{i : f_T(\mathbf{x}_i) = y_i\}|}$
\end{algorithmic}
\end{algorithm}

\section{Experimental Setup}
\label{sec:setup}

\subsection{Datasets}

We evaluate on five network intrusion detection benchmarks spanning two decades of traffic evolution (Table~\ref{tab:datasets}).
NSL-KDD~\cite{tavallaee2009detailed} serves as a legacy reference; core conclusions are independently validated on four modern datasets (2015--2022).
For CIC-IDS-2018, we select 25 features from the original 79 by removing timestamps, near-zero-variance columns, and highly correlated pairs; the DC/IC/UC partition applies to these 25 retained features.
RT-IoT2022 and HIKARI-2021 use all FlowMeter-derived features (83 and 82 respectively), with the forward/backward directionality providing a natural basis for the DC/UC partition.

\begin{table}[!t]
\centering
\caption{Dataset characteristics and controllability partitions. The table defines the feature spaces under which each transfer experiment is evaluated; it does not by itself imply relative vulnerability across datasets.}
\label{tab:datasets}
\small
\begin{tabular}{lcccc}
\toprule
\textbf{Dataset} & \textbf{Year} & \textbf{Features} & \textbf{Classes} & \textbf{DC/IC/UC} \\
\midrule
NSL-KDD~\cite{tavallaee2009detailed} & 1999 & 41 & 5 & 20/9/12 \\
UNSW-NB15~\cite{moustafa2015unsw} & 2015 & 41 & 10 & 15/11/15 \\
CIC-IDS-2018~\cite{sharafaldin2018toward} & 2018 & 25 & 4 & 14/7/4 \\
HIKARI-2021~\cite{ferriyan2021generating} & 2021 & 82 & 3 & 25/32/25 \\
RT-IoT2022~\cite{rtiot2022} & 2022 & 83 & 5 & 27/32/24 \\
\bottomrule
\end{tabular}
\end{table}

\subsection{Target Classifiers}

\textbf{LLM targets (7).}
Qwen3-8B, Qwen3-14B-FT, Qwen3-32B-FT, Phi3-Mini (3.8B), LLaMA3-8B, Mistral-7B, and Gemma2-9B.
Fine-tuned variants (suffix -FT) are instruction-tuned on in-domain flow descriptions via LoRA~\cite{hu2022lora}.
Because network intrusion benchmarks are typically class-imbalanced, we use clean macro-F1 rather than accuracy as the competence criterion~\cite{he2009learning}.
We retain only configurations achieving clean macro-F1 $\geq 0.45$, yielding 27 valid evaluation cells (Table~\ref{tab:valid_cells}).
This liberal threshold is deliberate: macro-F1~$= 0.45$ indicates above-random classification (random baseline $\approx 0.20$--$0.25$ for 4--5 classes), and excluding borderline cells would obscure how vulnerability correlates with competence.
To control for the concern that weak classifiers inflate ASR, we also report results restricted to F1~$\geq 0.60$ and~$\geq 0.70$ in Section~\ref{sec:f1_sensitivity}; conclusions hold at all thresholds.

\textbf{ML targets (2).}
LightGBM (tree ensemble) and DNN-IDS (3-layer MLP, 256--128--64 units).
We report comparisons disaggregated by ML target to avoid conflating the different surrogate-target transfer dynamics of tree$\to$tree versus tree$\to$neural architectures.

\textbf{Surrogate.}
XGBoost trained on the same data partition, used exclusively for adversarial example generation.

\begin{table}[!t]
\centering
\caption{Valid LLM evaluation cells (clean macro-F1 $\geq$ 0.45).}
\label{tab:valid_cells}
\small
\begin{tabular}{lcc}
\toprule
\textbf{Dataset} & \textbf{\# Valid} & \textbf{F1 range} \\
\midrule
NSL-KDD & 5 & 0.54--0.68 \\
UNSW-NB15 & 4 & 0.47--0.75 \\
CIC-IDS-2018 & 5 & 0.45--0.85 \\
RT-IoT2022 & 7 & 0.63--0.96 \\
HIKARI-2021 & 6 & 0.45--0.99 \\
\midrule
\textbf{Total} & \textbf{27} & \\
\bottomrule
\end{tabular}
\end{table}

\subsection{LLM Classification Pipeline}
\label{sec:pipeline}

Flow features are serialized into structured natural language prompts via a per-dataset template.
This pipeline is the experimental condition that distinguishes LLM targets from tabular ML targets: adversarial perturbations are generated in normalized feature space, but LLM inference occurs after text serialization.
For NSL-KDD, a representative prompt is:

{\small
\begin{verbatim}
Classify this network flow as: Normal,
DoS, Probe, CredentialAccess, or
Exploitation.

Flow: Protocol TCP to 'http' (flag: SF,
SYN then FIN). Duration 30s, 1024B out,
512B in. Login active. Connections: 5 to
same host (SYN-err 0%.)
\end{verbatim}
}

\noindent Numeric features are formatted as integers where semantically appropriate (\eg, byte counts, durations) and as percentages for rate features.
Feature names are embedded in the prompt to provide semantic context.

\textbf{Inference.}
We use vLLM~\cite{kwon2023efficient} with greedy decoding (temperature~$= 0$), max new tokens~$= 16$, batch size~$= 32$.
The model's text output is parsed against the label vocabulary; outputs not matching any label are treated as misclassifications.

\textbf{Fine-tuning.}
LoRA adapters~\cite{hu2022lora} (rank~$= 16$, $\alpha = 32$) are trained on the training split of each dataset for 3 epochs with learning rate $2\times10^{-4}$.
Non-fine-tuned models (Gemma2, Phi3, LLaMA3, Mistral) use the base instruction-tuned checkpoints with the same prompt format.

\subsection{Attack Configuration}

Perturbation budgets: $\varepsilon \in \{0.01, 0.05, 0.15, 0.30, 0.50\}$ in $L_\infty$-normalized feature space (features min-max scaled to $[0,1]$ using training-set statistics).
An $\varepsilon = 0.05$ budget therefore corresponds to a raw delta of $5\%$ of each feature's training-set range: for compact binary or flag features (original range $[0,1]$) this is a raw change of 0.05, while for high-volume byte-count features (range $\sim$$10^5$) this corresponds to changes of several thousand bytes.
We note that $\varepsilon \leq 0.15$ maps to raw changes that are plausibly achievable through application-layer traffic shaping for most flow statistics; $\varepsilon = 0.30$--$0.50$ should be treated as upper-bound exploration rather than operationally realistic budgets.
Each configuration is repeated across 5 random seeds (42, 123, 2025, 7, 99), which control the evaluation-set subsampling and attack initialization (random start in $\varepsilon$-ball); the LLM target and surrogate weights are fixed.
Evaluation uses 500 stratified samples per class.
Results are reported as mean~$\pm$~std across all ($\varepsilon$, seed) combinations unless otherwise noted.
We note that ($\varepsilon$, seed) pairs sharing the same evaluation subset are not fully independent; accordingly, the variance estimates should be interpreted as measuring sensitivity to perturbation budget and initialization, not as independent replications.
For statistical analysis, we report within-dataset tests over (model, $\varepsilon$, seed) tuples as descriptive evidence and use a dataset-level paired analysis ($n = 5$ datasets) as the primary guard against inflated significance from within-dataset pooling (Section~\ref{sec:significance}).
This setup lets the budget analysis test whether the LLM-vs-ML pattern is confined to one perturbation regime or persists across small and large changes.

\subsection{Attack Pipeline}
\label{sec:attack_pipeline}

For LLM targets, the end-to-end pipeline is:
(i)~raw flow features (numeric, integer, and binary) are min-max normalized to $[0,1]$ using training-set statistics;
(ii)~the adversarial perturbation is applied in normalized space, restricted to DC feature indices and clipped to $[\max(0,\,x_i-\varepsilon),\,\min(1,\,x_i+\varepsilon)]$;
(iii)~the perturbed normalized vector is serialized as a structured text prompt (feature name: value pairs) and passed to the LLM target.
For integer-valued DC features (\eg, packet counts, duration), perturbations are applied in normalized space and retain floating-point precision in the LLM prompt.
A validity audit on NSL-KDD (17 integer-valued DC features) compares ASR with and without post-perturbation rounding: rounding has no measurable effect at $\varepsilon = 0.01$, reduces ASR by 3.6\% at $\varepsilon = 0.05$, and by 22--26\% at $\varepsilon \geq 0.15$, indicating ASR is conservative at low budgets and may be modestly overestimated at $\varepsilon \geq 0.15$.
For DC features with large raw ranges (e.g., \texttt{fwd\_pkts\_tot}), since DC features are \emph{directly controllable}, the attacker can set any valid integer, incurring at most one rounding step.
The LLM-vs-LGB ordering is unchanged at $\varepsilon=0.05$ vs.\ $\varepsilon=0.15$ across all four modern datasets (RT-IoT2022: $8.3\times$/$7.2\times$; CIC-IDS-2018: $1.48\times$/$1.45\times$; HIKARI-2021: $0.81\times$/$0.80\times$; UNSW-NB15: $0.97\times$/$0.93\times$), confirming that main conclusions hold at operational budgets.
The surrogate $f_S$ (XGBoost by default) operates directly on the normalized feature vector and is never queried through the text serialization layer.

\subsection{Metrics}

\textbf{Attack Success Rate (ASR):} Fraction of correctly classified clean samples that are misclassified after perturbation.
This metric includes all traffic classes (normal and attack classes).

\textbf{Evasion ASR:} ASR restricted to attack-class samples only, excluding benign/Normal traffic.
Evasion ASR is the security-operational metric: it measures how frequently an ongoing attack is misclassified after adversarial modification, regardless of how the model handles benign flows.
Evasion ASR values for transfer attacks (mean across LLM configurations, clean F1~$\geq 0.45$) are:
NSL-KDD~0.39, UNSW-NB15~0.41, CIC-IDS-2018~0.78, RT-IoT2022~0.64, HIKARI-2021~0.57.
All values are computed from per-class ASR logs under the DC\,=\,27/25 taxonomy.
The relative ordering across datasets and the LLM-vs-ML gap are consistent under both Evasion ASR and the standard ASR metric.

\textbf{Constraint Violation Rate (CVR):} Fraction of adversarial examples with any non-DC feature modified---zero by construction, verified as implementation correctness.
\textbf{Clean Macro-F1:} Baseline classification performance without attack.

\section{Results}
\label{sec:results}

\subsection{LLM Attack Success Rates}

Table~\ref{tab:llm_results} presents ASR for all 27 valid LLM configurations.
FD-PGD and NES achieve nontrivial success rates across all datasets (NSL-KDD mean 0.286 to CIC-IDS-2018 mean 0.691), while greedy shows near-zero LLM transferability (RT-IoT2022 mean 0.005; HIKARI-2021 mean 0.027).
LLM-vs-ML comparison requires Tables~\ref{tab:ml_results} and~\ref{tab:comparison}.

\begin{table*}[!t]
\centering
\caption{LLM attack success rates (mean $\pm$ std across $\varepsilon$ and seeds). Only valid cells with clean macro-F1 $\geq 0.45$ are reported; bold indicates the highest ASR per row.}
\label{tab:llm_results}
\small
\begin{tabular}{ll ccc}
\toprule
\textbf{Dataset} & \textbf{Model (Clean F1)} & \textbf{FD-PGD Transfer} & \textbf{Greedy} & \textbf{NES} \\
\midrule
\multirow{5}{*}{NSL-KDD}
 & Gemma2-9B (0.54) & 0.290$\pm$0.019 & 0.155$\pm$0.016 & \textbf{0.286$\pm$0.029} \\
 & LLaMA3-8B (0.68) & 0.332$\pm$0.021 & 0.134$\pm$0.012 & \textbf{0.345$\pm$0.021} \\
 & Phi3-Mini (0.63) & 0.192$\pm$0.018 & 0.061$\pm$0.014 & \textbf{0.215$\pm$0.027} \\
 & Qwen3-14B-FT (0.66) & \textbf{0.293$\pm$0.015} & 0.135$\pm$0.010 & 0.284$\pm$0.016 \\
 & Qwen3-32B-FT (0.68) & 0.324$\pm$0.033 & 0.081$\pm$0.014 & \textbf{0.389$\pm$0.026} \\
\midrule
\multirow{4}{*}{UNSW-NB15}
 & Gemma2-9B (0.56) & 0.208$\pm$0.101 & 0.015$\pm$0.005 & \textbf{0.324$\pm$0.122} \\
 & LLaMA3-8B (0.61) & 0.265$\pm$0.070 & 0.041$\pm$0.028 & \textbf{0.266$\pm$0.052} \\
 & Mistral-7B (0.47) & \textbf{0.386$\pm$0.033} & 0.011$\pm$0.003 & 0.383$\pm$0.065 \\
 & Qwen3-8B (0.75) & 0.312$\pm$0.050 & 0.040$\pm$0.005 & \textbf{0.370$\pm$0.072} \\
\midrule
\multirow{5}{*}{CIC-IDS-2018}
 & LLaMA3-8B (0.45) & 0.515$\pm$0.021 & 0.262$\pm$0.021 & \textbf{0.541$\pm$0.054} \\
 & Mistral-7B (0.73) & \textbf{0.723$\pm$0.020} & 0.072$\pm$0.020 & 0.717$\pm$0.031 \\
 & Qwen3-14B-FT (0.75) & \textbf{0.773$\pm$0.020} & 0.574$\pm$0.028 & 0.733$\pm$0.047 \\
 & Qwen3-32B-FT (0.82) & \textbf{0.742$\pm$0.013} & 0.577$\pm$0.025 & 0.696$\pm$0.050 \\
 & Qwen3-8B (0.85) & \textbf{0.702$\pm$0.028} & 0.407$\pm$0.061 & 0.666$\pm$0.048 \\
\midrule
\multirow{7}{*}{RT-IoT2022}
 & Gemma2-9B (0.96) & 0.498$\pm$0.034 & 0.003$\pm$0.001 & \textbf{0.522$\pm$0.034} \\
 & LLaMA3-8B (0.63) & 0.711$\pm$0.021 & 0.005$\pm$0.002 & \textbf{0.735$\pm$0.037} \\
 & Mistral-7B (0.74) & 0.424$\pm$0.064 & 0.007$\pm$0.002 & \textbf{0.428$\pm$0.062} \\
 & Phi3-Mini (0.89) & 0.503$\pm$0.077 & 0.005$\pm$0.001 & \textbf{0.527$\pm$0.055} \\
 & Qwen3-14B-FT (0.93) & \textbf{0.589$\pm$0.051} & 0.006$\pm$0.002 & 0.548$\pm$0.017 \\
 & Qwen3-32B-FT (0.67) & \textbf{0.473$\pm$0.038} & 0.005$\pm$0.002 & 0.472$\pm$0.031 \\
 & Qwen3-8B (0.88) & \textbf{0.581$\pm$0.042} & 0.003$\pm$0.002 & 0.536$\pm$0.027 \\
\midrule
\multirow{6}{*}{HIKARI-2021}
 & Gemma2-9B (0.45)$^\ddagger$ & \textbf{0.110$\pm$0.011} & 0.019$\pm$0.004 & 0.102$\pm$0.008 \\
 & LLaMA3-8B (0.99) & \textbf{0.458$\pm$0.079} & 0.108$\pm$0.044 & 0.436$\pm$0.057 \\
 & Mistral-7B (0.99) & 0.345$\pm$0.010 & 0.023$\pm$0.023 & \textbf{0.349$\pm$0.008} \\
 & Phi3-Mini (0.74) & \textbf{0.303$\pm$0.074} & 0.001$\pm$0.001 & 0.302$\pm$0.016 \\
 & Qwen3-14B-FT (0.99) & 0.294$\pm$0.025 & 0.007$\pm$0.004 & \textbf{0.319$\pm$0.010} \\
 & Qwen3-8B (0.99) & \textbf{0.340$\pm$0.048} & 0.005$\pm$0.006 & 0.332$\pm$0.005 \\
\midrule
\multicolumn{5}{l}{\footnotesize $^\ddagger$Gemma2-9B/HIKARI F1 = 0.45 (inclusion threshold); low ASR reflects limited classification competence on this 3-class dataset.}\\
\bottomrule
\end{tabular}
\end{table*}

\subsection{ML Baseline Attack Success Rates}

Table~\ref{tab:ml_results} presents ASR for the two ML targets under the identical black-box transfer protocol.
LightGBM, as a fellow tree ensemble, shows moderate transfer susceptibility across datasets.
DNN-IDS exhibits near-zero ASR on most datasets (0.008--0.064 for transfer), consistent with the large architectural gap between a tree surrogate and a neural-network target; this asymmetry is central to the disaggregated comparison below.

\begin{table}[!t]
\centering
\caption{ML target ASR under black-box transfer using an XGBoost surrogate, reported separately for tree and neural targets.}
\label{tab:ml_results}
\scriptsize
\setlength{\tabcolsep}{2.0pt}
\renewcommand{\arraystretch}{1.08}
\begin{tabular}{@{}p{0.22\columnwidth}p{0.15\columnwidth}ccc@{}}
\toprule
\textbf{Dataset} & \textbf{Target} & \textbf{Transfer} & \textbf{Greedy} & \textbf{NES} \\
\midrule
\multirow{2}{*}{NSL-KDD}
 & LightGBM & 0.320$\pm$0.091 & 0.202$\pm$0.019 & 0.403$\pm$0.066 \\
 & DNN-IDS & 0.009$\pm$0.008 & 0.002$\pm$0.003 & 0.014$\pm$0.010 \\
\midrule
\multirow{2}{*}{UNSW-NB15}
 & LightGBM & 0.327$\pm$0.103 & 0.095$\pm$0.022 & 0.512$\pm$0.083 \\
 & DNN-IDS & 0.064$\pm$0.052 & 0.030$\pm$0.024 & 0.096$\pm$0.067 \\
\midrule
\multirow{2}{*}{CIC-IDS-2018}
 & LightGBM & 0.478$\pm$0.054 & 0.631$\pm$0.030 & 0.628$\pm$0.072 \\
 & DNN-IDS & 0.222$\pm$0.164 & 0.186$\pm$0.170 & 0.328$\pm$0.198 \\
\midrule
\multirow{2}{*}{RT-IoT2022}
 & LightGBM & 0.075$\pm$0.036 & 0.007$\pm$0.004 & 0.208$\pm$0.081 \\
 & DNN-IDS & 0.014$\pm$0.018 & 0.000$\pm$0.000 & 0.021$\pm$0.024 \\
\midrule
\multirow{2}{*}{HIKARI-2021}
 & LightGBM & 0.441$\pm$0.062 & 0.189$\pm$0.005 & 0.598$\pm$0.064 \\
 & DNN-IDS & 0.039$\pm$0.061 & 0.000$\pm$0.000 & 0.052$\pm$0.064 \\
\bottomrule
\end{tabular}
\end{table}

\subsection{LLM vs.\ ML Vulnerability Comparison}

Table~\ref{tab:comparison} disaggregates the comparison by ML target type.
LLM~vs.~LightGBM and LLM~vs.~DNN-IDS are reported separately, since averaging the two ML targets would obscure the distinct transfer dynamics of tree$\to$tree versus tree$\to$neural pathways.
This table is the primary evidence for the central thesis: LLMs are not uniformly more vulnerable than a strong tabular comparator.

\begin{table}[!t]
\centering
\caption{LLM vs.\ ML transfer vulnerability disaggregated by ML target. $p$-values are within-dataset one-sided Mann--Whitney~U tests over pooled ASR tuples; dataset-level statistics are reported in Section~\ref{sec:significance}.}
\label{tab:comparison}
\small
\begin{tabular}{l ccc cc}
\toprule
& \textbf{LLM} & \textbf{LGB} & \textbf{DNN} & \textbf{LLM/} & $p$ \\
\textbf{Dataset} & \textbf{Avg} & & & \textbf{LGB} & (vs LGB) \\
\midrule
NSL-KDD & 0.286 & 0.320 & 0.009 & 0.9$\times^\dagger$ & 0.98 \\
UNSW-NB15 & 0.293 & 0.327 & 0.064 & 0.9$\times^\dagger$ & 0.90 \\
CIC-2018 & 0.691 & 0.478 & 0.222 & 1.4$\times$ & $<$0.001 \\
RT-IoT & 0.540 & 0.075 & 0.014 & 7.2$\times$ & $<$0.001 \\
HIKARI & 0.308 & 0.441 & 0.039 & 0.70$\times^\ddagger$ & $>$0.99 \\
\bottomrule
\end{tabular}
\end{table}

\begin{figure}[!t]
\centering
\includegraphics[width=\columnwidth]{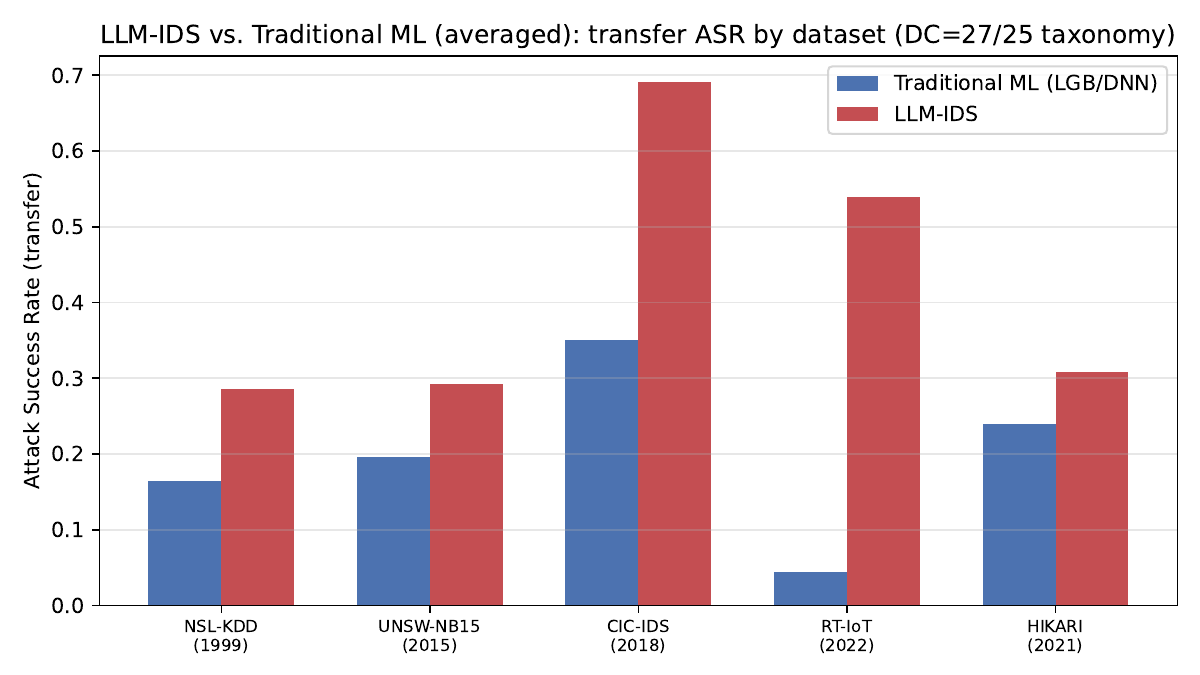}
\caption{LLM vs.\ ML attack success rates under transfer attack (averaged ML reference). Disaggregated LightGBM comparison is in Table~\ref{tab:comparison}.}
\label{fig:llm_vs_ml}
\end{figure}

The disaggregated analysis reveals that the LLM-vs-LGB gap depends on both dataset and perturbation regime.
On CIC-IDS-2018 and RT-IoT2022, LLMs are more vulnerable than LightGBM, with ratios of $1.4\times$ and $7.2\times$ respectively.
On the two legacy datasets (NSL-KDD, UNSW-NB15), LLM and LightGBM vulnerability are comparable.
HIKARI-2021 presents a qualitatively distinct result under the DC\,=\,25 taxonomy applied consistently to both LLM and ML targets.
When LightGBM is also constrained to the 25 DC features (see Section~\ref{sec:hikari_crossover}), its transfer ASR rises to 0.441---above the LLM average of 0.308 at every tested $\varepsilon$ (LGB: 0.347--0.526, LLM: 0.248--0.352; within-dataset MW test LGB\,$>$\,LLM: $p = 3 \times 10^{-6}$, Cohen's $d = 1.45$).
The LLM/LGB ratio is 0.70$\times$ overall, and LLaMA3-8B---the highest-performing LLM on HIKARI (F1 = 0.99)---also remains below LightGBM at all $\varepsilon \geq 0.15$.
One possible explanation is that the 25 fwd\_-prefixed DC features on HIKARI align more closely with LightGBM's split boundaries than with attention-weighted token representations: axis-aligned perturbations in these features can cross multiple tree thresholds simultaneously, while the LLM may redistribute attention across the full constrained feature sequence.
This explanation is a hypothesis about the observed reversal, not an experimentally isolated cause.

The DNN-IDS comparison is confounded by the surrogate-target architectural mismatch: the XGBoost surrogate transfers poorly to DNN-IDS (ASR~$= 0.008$--$0.064$ outside CIC), likely reflecting poor tree$\to$neural transfer rather than inherent DNN resistance.
We therefore base our primary vulnerability conclusions on the LLM-vs-LightGBM comparison, which controls for this confound.

\textbf{Summary.}
Under tree-surrogate transfer, LLMs are more vulnerable than LightGBM on CIC-IDS-2018 ($1.4\times$) and RT-IoT2022 ($7.2\times$); on HIKARI-2021, LightGBM is more vulnerable ($0.70\times$ LLM/LGB, DC\,=\,25); and on NSL-KDD and UNSW-NB15, susceptibility is comparable.
When pooling LightGBM and DNN-IDS as the ML baseline, LLMs have higher ASR on all five datasets, but this aggregate is partly driven by weak transfer to DNN-IDS and is therefore secondary to the disaggregated LightGBM comparison.

\section{Analysis}
\label{sec:analysis}

The Analysis section tests whether the main comparison is an artifact of one attack family, one surrogate, one perturbation budget, or weak LLM classifiers.

\subsection{Validation of the Transfer Hierarchy}

\begin{figure}[!t]
\centering
\includegraphics[width=\columnwidth]{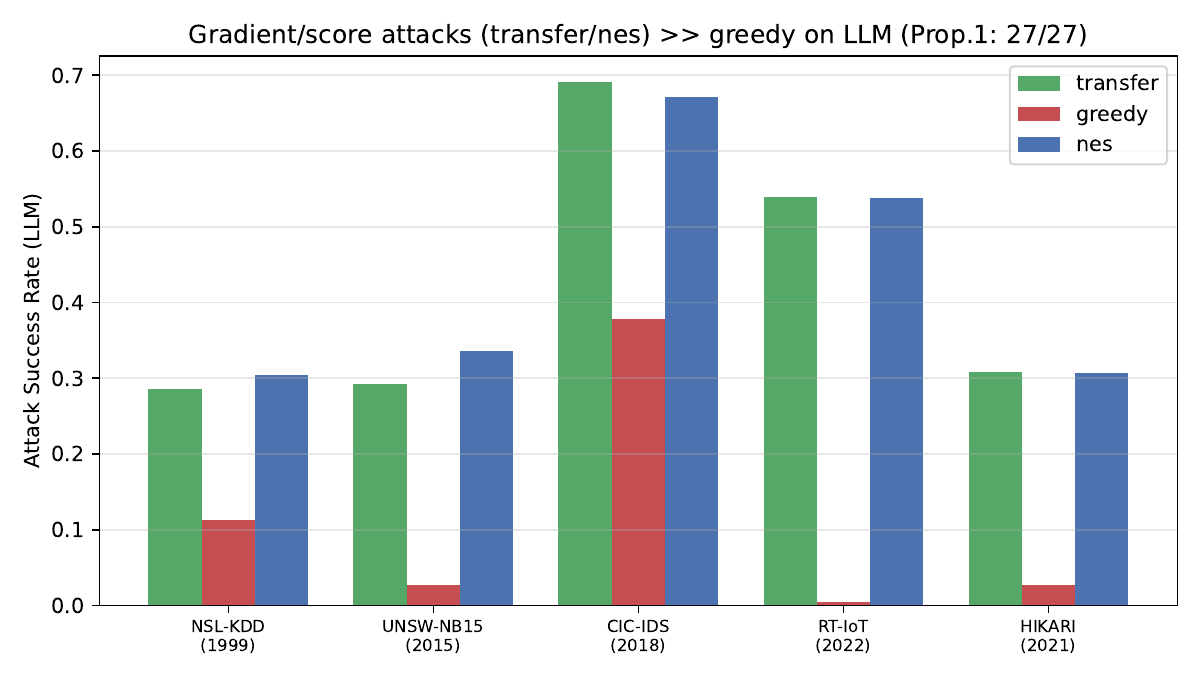}
\caption{Transfer hierarchy: FD-PGD and NES dominate greedy on all 27 LLM cells and 9/10 ML cells. The CIC/LightGBM exception ($\star$) corresponds to near-identical surrogate-target agreement ($\rho = 0.98$).}
\label{fig:attack_comparison}
\end{figure}

Fig.~\ref{fig:attack_comparison} visualizes Empirical Finding~1.
The figure asks whether the lower transfer of greedy attacks is specific to LLMs or also appears when the target architecture differs from the surrogate in ML baselines.
On LLMs, the hierarchy holds in all 27 valid cells: both transfer and NES attacks strictly outperform greedy.
The margin is large: dataset-averaged greedy ASR on LLMs ranges from 0.005 (RT-IoT2022) to 0.378 (CIC-IDS-2018), while transfer ASR ranges from 0.286 to 0.691.

On ML targets, the hierarchy holds in 9 of 10 cells under the same black-box protocol (Table~\ref{tab:prop1_ml}).
The table supports the interpretation that transfer behavior depends on surrogate-target similarity, not only on whether the target is an LLM.

\begin{table}[!t]
\centering
\caption{Transfer hierarchy on ML targets. T/G/N = transfer/greedy/NES mean ASR; $\rho$ = XGBoost--target prediction agreement.}
\label{tab:prop1_ml}
\small
\begin{tabular}{llccccc}
\toprule
\textbf{Dataset} & \textbf{Target} & \textbf{T} & \textbf{G} & \textbf{N} & $\rho$ & \textbf{Holds?} \\
\midrule
NSL-KDD & LGB & .32 & .20 & .40 & .89 & $\checkmark$ \\
NSL-KDD & DNN & .01 & .00 & .01 & .82 & $\checkmark$ \\
UNSW & LGB & .33 & .10 & .51 & .91 & $\checkmark$ \\
UNSW & DNN & .06 & .03 & .10 & .85 & $\checkmark$ \\
CIC-2018 & LGB & .48 & .63 & .63 & \textbf{.98} & $\times$ \\
CIC-2018 & DNN & .22 & .19 & .33 & .88 & $\checkmark$ \\
RT-IoT & LGB & .08 & .01 & .21 & .93 & $\checkmark$ \\
RT-IoT & DNN & .01 & .00 & .02 & .84 & $\checkmark$ \\
HIKARI & LGB & .44 & .19 & .60 & .90 & $\checkmark$ \\
HIKARI & DNN & .04 & .00 & .05 & .81 & $\checkmark$ \\
\bottomrule
\end{tabular}
\end{table}

\subsection{Mechanistic Explanation: Architectural Similarity}
\label{sec:mechanism}

The single exception---CIC-IDS-2018/LightGBM, where greedy (0.631) exceeds transfer (0.478)---is consistent with the architectural similarity mechanism.

On CIC-IDS-2018, the XGBoost surrogate and LightGBM target achieve prediction agreement $\rho = 0.98$---functionally near-identical classifiers.
Under such minimal dissimilarity, greedy perturbations optimized on the surrogate transfer nearly losslessly.
In contrast, XGBoost-to-DNN-IDS agreement on the same dataset is $\rho = 0.88$, and greedy ASR drops to 0.186---restoring the hierarchy.

The $\rho$ values span both extremes:
\begin{itemize}
    \item $\rho \approx 1.0$ (XGBoost $\to$ LightGBM on CIC): greedy transfers with high ASR, hierarchy absent.
    \item $\rho \ll 1.0$ (XGBoost $\to$ LLM, any dataset): greedy transfer collapses to near-zero, hierarchy maximal.
\end{itemize}

More broadly, $\rho$ values across all target cells span the range $[0.25, 0.98]$.
LLM targets cluster at $\rho \in [0.25, 0.68]$, reflecting the fundamental architectural gap between tree-ensemble surrogates and Transformer-based classifiers.
ML targets occupy $\rho \in [0.85, 0.98]$ (LightGBM) and $\rho \in [0.72, 0.88]$ (DNN-IDS).
The transfer hierarchy (gradient/score $>$ greedy) holds universally for $\rho < 0.95$ and breaks only at near-identity similarity ($\rho = 0.98$), consistent with a monotone relationship between architectural dissimilarity and the transferability advantage of multi-coordinate perturbations.
This analysis should not be overread as proving causality from $\rho$ alone; $\rho$ is a proxy for architectural similarity and is used here to interpret the observed exception.

\subsection{Cross-Surrogate Transfer Check}
\label{sec:surrogate}

\begin{figure}[!t]
\centering
\includegraphics[width=\columnwidth]{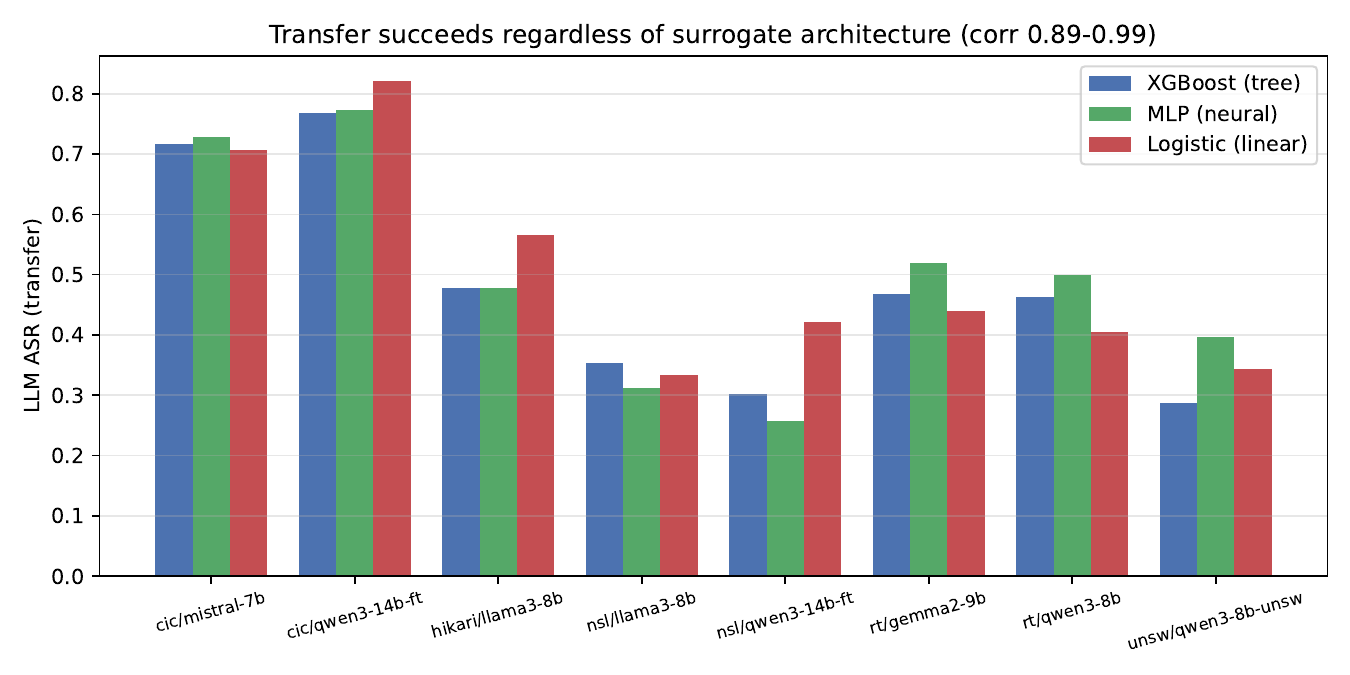}
\caption{Cross-surrogate transfer: XGBoost, MLP, and logistic regression surrogates produce comparable mean ASR across 8 LLM cells.}
\label{fig:surrogate}
\end{figure}

To test whether findings depend on the XGBoost surrogate, we evaluate three architecturally diverse surrogates---XGBoost, MLP, and logistic regression---on 8~LLM cells across all five datasets (Table~\ref{tab:surrogate}, Fig.~\ref{fig:surrogate}).
This analysis answers a reproducibility question: whether transfer to LLM targets disappears when adversarial examples are generated by a non-tree surrogate.

\begin{table}[!t]
\centering
\caption{Multi-surrogate transfer ASR to LLM targets (FD-PGD, mean across $\varepsilon \in \{0.05, 0.15, 0.30\}$, 2 seeds). $\ddagger$HIKARI uses DC\,=\,25 taxonomy.}
\label{tab:surrogate}
\small
\begin{tabular}{ll ccc}
\toprule
\textbf{Dataset} & \textbf{LLM Target} & \textbf{XGB} & \textbf{MLP} & \textbf{Logistic} \\
\midrule
NSL-KDD & LLaMA3-8B & 0.354 & 0.312 & 0.334 \\
NSL-KDD & Qwen3-14B-FT & 0.303 & 0.257 & 0.422 \\
CIC-2018 & Mistral-7B & 0.717 & 0.729 & 0.706 \\
CIC-2018 & Qwen3-14B-FT & 0.768 & 0.774 & 0.821 \\
RT-IoT & Gemma2-9B & 0.506 & 0.520 & 0.440 \\
RT-IoT & Qwen3-8B & 0.594 & 0.499 & 0.404 \\
HIKARI & LLaMA3-8B$^\ddagger$ & 0.479 & 0.478 & 0.566 \\
UNSW-NB15 & Qwen3-8B-UNSW & 0.287 & 0.396 & 0.343 \\
\midrule
\textbf{Mean (8 cells)} & & \textbf{0.501} & \textbf{0.496} & \textbf{0.505} \\
\bottomrule
\end{tabular}
\end{table}

All three surrogates produce nontrivial adversarial transfer against every tested LLM target, with cross-surrogate standard deviation within each cell of 0.042 on average.
Across all eight cells spanning all five datasets, the linear logistic-regression surrogate achieves mean ASR comparable to XGBoost (0.505 vs.\ 0.501).
This result does not prove a mechanism, but it weakens the concern that the observed transfer is an artifact of one particular XGBoost surrogate.
The HIKARI-2021 cell (LLaMA3-8B, DC\,=\,25 taxonomy) checks the finding on a dataset where LightGBM is more vulnerable than the LLM average under tree-surrogate transfer: all three surrogate architectures produce nontrivial adversarial examples (XGB\,=\,0.479, MLP\,=\,0.478, Logistic\,=\,0.566).
The UNSW-NB15 cell (Qwen3-8B-UNSW) shows lower absolute ASR (XGB\,=\,0.287, MLP\,=\,0.396, Logistic\,=\,0.343), consistent with the lower overall LLM vulnerability on this dataset (Table~\ref{tab:comparison}).

\subsection{Perturbation Budget Analysis}

\begin{figure}[!t]
\centering
\includegraphics[width=\columnwidth]{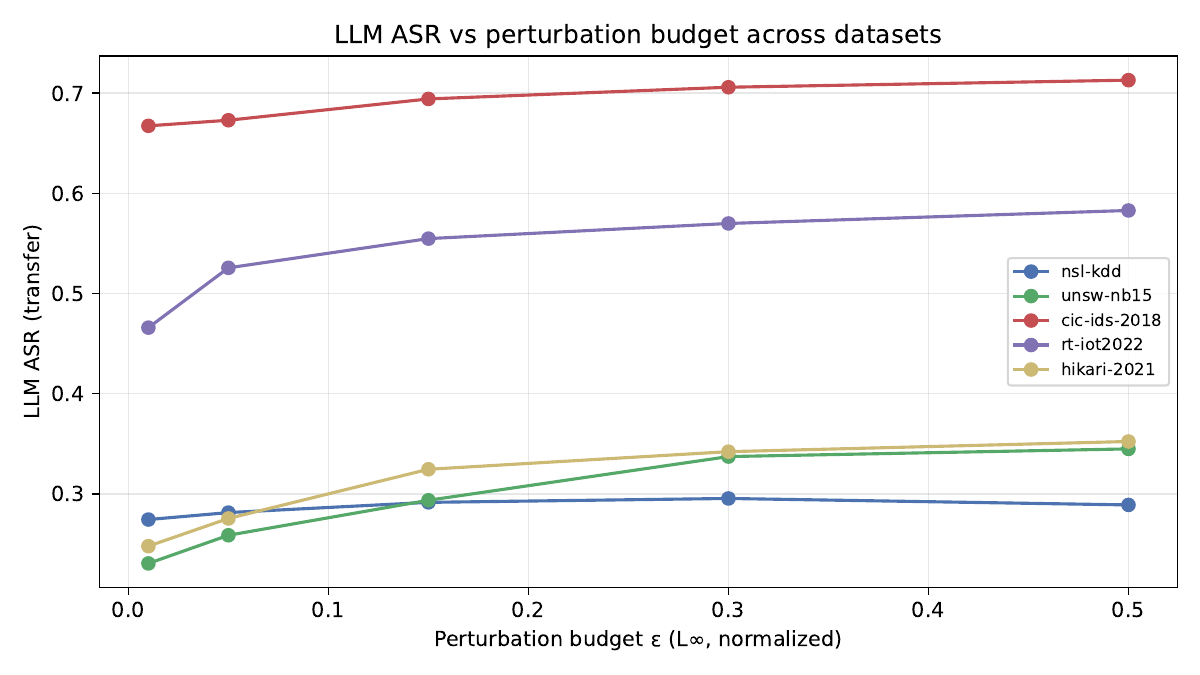}
\caption{Transfer ASR vs.\ $\varepsilon$ (averaged ML\,=\,LightGBM\,+\,DNN-IDS). The averaged LLM-vs-ML gap persists across budgets; against LightGBM alone, HIKARI-2021 reverses (Section~\ref{sec:hikari_crossover}).}
\label{fig:eps}
\end{figure}

Fig.~\ref{fig:eps} and Table~\ref{tab:per_eps} show how ASR varies with $\varepsilon$.
This analysis asks whether the central comparison is an artifact of a single perturbation budget.
On RT-IoT2022, the LLM vulnerability advantage over ML holds at every perturbation level, with the ratio peaking at $\varepsilon = 0.01$ (LLM/ML-avg: $52\times$) and decreasing as $\varepsilon$ grows because ML ASR increases faster with budget; LLM ASR saturates beyond $\varepsilon = 0.15$--$0.30$.
On HIKARI-2021, LLMs exceed the averaged ML baseline at every $\varepsilon$ (LLM: 0.248--0.352 vs.\ ML-avg: 0.174--0.338), though the gap is narrow at $\varepsilon = 0.50$ (0.352 vs.\ 0.338); when compared against LightGBM alone, LightGBM exceeds LLMs at all $\varepsilon$---consistent with the per-$\varepsilon$ analysis in Section~\ref{sec:hikari_crossover}.

\begin{table}[!t]
\centering
\caption{Per-$\varepsilon$ LLM vs.\ ML mean transfer ASR (averaged ML baseline).}
\label{tab:per_eps}
\small
\begin{tabular}{l cc cc cc}
\toprule
& \multicolumn{2}{c}{$\varepsilon=0.01$} & \multicolumn{2}{c}{$\varepsilon=0.15$} & \multicolumn{2}{c}{$\varepsilon=0.50$} \\
\cmidrule(lr){2-3}\cmidrule(lr){4-5}\cmidrule(lr){6-7}
\textbf{Dataset} & LLM & ML & LLM & ML & LLM & ML \\
\midrule
NSL-KDD & .274 & .087 & .291 & .181 & .289 & .219 \\
UNSW & .230 & .090 & .293 & .188 & .345 & .275 \\
CIC-2018 & .709 & .202 & .739 & .354 & .756 & .501 \\
RT-IoT & .466 & .009 & .555 & .040 & .583 & .085 \\
HIKARI & .248 & .174 & .324 & .225 & .352 & .338 \\
\bottomrule
\end{tabular}
\end{table}

\subsection{Classifier Competence Sensitivity}
\label{sec:f1_sensitivity}

To control for the concern that attacking weak classifiers (low clean~F1) inflates ASR, we repeat the LLM-vs-ML comparison at F1 thresholds of 0.60 and 0.70.
At F1~$\geq 0.60$ (22 cells), LLMs remain more vulnerable than the averaged ML baseline on all five datasets.
At F1~$\geq 0.70$ (15 cells), the comparison is available on four datasets (NSL-KDD drops out as no LLM exceeds F1~$= 0.70$); versus the averaged ML baseline, LLMs remain more vulnerable on all four.
The advantage over LightGBM specifically is maintained at F1~$\geq 0.70$ on RT-IoT2022 ($> 7\times$) and CIC-IDS-2018, which reduces the concern that the vulnerability is only an artifact of attacking weak classifiers.

\subsection{Statistical Significance}
\label{sec:significance}

Table~\ref{tab:stats} reports one-sided Mann-Whitney~U tests for the LLM-vs-ML~(averaged) transfer ASR comparison within each dataset.
All five within-dataset tests yield $p < 0.01$ with large effect sizes (Cohen's $d \geq 0.75$), with four of five reaching $p < 0.001$.
Because these tests pool (model, $\varepsilon$, seed) tuples, they should not be read as five independent confirmations of the same global claim.
The disaggregated LLM-vs-LightGBM comparison (Table~\ref{tab:comparison}) shows the central pattern more directly: LLMs are higher on CIC-IDS-2018 and RT-IoT2022, comparable on NSL-KDD and UNSW-NB15, and lower on HIKARI-2021.

\textbf{Dataset-level check.}
To guard against within-dataset dependence inflating the per-dataset tests, we also performed a dataset-level paired analysis using per-dataset LLM and ML means as observations ($n = 5$ paired datasets, one observation per dataset).
A Wilcoxon signed-rank test yields $W = 15$ ($p = 0.031$, one-sided), Cohen's $d = 1.21$ at the dataset level, supporting the claim that LLMs show higher ASR than the averaged ML baseline consistently across all five datasets.
Bootstrap 95\% confidence intervals computed by resampling over models within each dataset (10{,}000 iterations) show CIC-IDS-2018 LLM mean $[0.598, 0.752]$ and RT-IoT2022 $[0.480, 0.610]$, both entirely above LightGBM (0.478 and 0.075 respectively); NSL-KDD and UNSW-NB15 CIs overlap with LightGBM, consistent with the ``comparable'' characterization, and HIKARI-2021 CI $[0.219, 0.386]$ lies below LightGBM (0.441), supporting the reversed-gap interpretation.

\begin{table}[!t]
\centering
\caption{Within-dataset one-sided Mann-Whitney~U tests, LLM vs.\ averaged-ML transfer ASR. Results pool (model, $\varepsilon$, seed) tuples; dataset-level paired result (Wilcoxon $W=15$, $p=0.031$, $d=1.21$) is reported in Section~\ref{sec:significance}.}
\label{tab:stats}
\small
\begin{tabular}{lcccc}
\toprule
\textbf{Dataset} & $U$ & $p$ & Cohen's $d$ & Sig. \\
\midrule
NSL-KDD & 4278 & $7.0 \times 10^{-5}$ & 0.97 & *** \\
UNSW-NB15 & 3468 & $5.7 \times 10^{-5}$ & 0.76 & *** \\
CIC-IDS-2018 & 5000 & $1.1 \times 10^{-23}$ & 3.02 & *** \\
RT-IoT2022 & 8750 & $2.2 \times 10^{-27}$ & 6.39 & *** \\
HIKARI-2021 & 542 & $2.4 \times 10^{-3}$ & 0.75 & ** \\
\bottomrule
\end{tabular}
\end{table}

\subsection{HIKARI: LGB Higher Vulnerability Under Corrected Taxonomy}
\label{sec:hikari_crossover}

When LightGBM is evaluated under the same DC\,=\,25 taxonomy used for LLMs, the per-$\varepsilon$ breakdown (Table~\ref{tab:hikari_crossover}) shows higher LightGBM ASR at all tested budgets.
This experiment is the clearest boundary case for the central thesis: it shows that LLMs are not uniformly more vulnerable than a strong tree comparator under the same controllability constraint.

\begin{table}[!t]
\centering
\caption{HIKARI-2021 transfer ASR per budget (DC\,=\,25 taxonomy). LightGBM exceeds the LLM average at every $\varepsilon$.}
\label{tab:hikari_crossover}
\footnotesize
\setlength{\tabcolsep}{4.0pt}
\renewcommand{\arraystretch}{1.08}
\begin{tabular}{@{}lccccc@{}}
\toprule
\textbf{Target} & \multicolumn{5}{c}{$\boldsymbol{\varepsilon}$} \\
\cmidrule(lr){2-6}
& 0.01 & 0.05 & 0.15 & 0.30 & 0.50 \\
\midrule
LLM avg & 0.248 & 0.275 & 0.324 & 0.342 & 0.352 \\
LightGBM & 0.347 & 0.406 & 0.448 & 0.480 & 0.526 \\
\textbf{LLM/LGB} & 0.71$\times$ & 0.68$\times$ & 0.72$\times$ & 0.71$\times$ & 0.67$\times$ \\
\bottomrule
\end{tabular}
\end{table}

LightGBM ASR grows monotonically from 0.347 to 0.526 (+52\%) as $\varepsilon$ increases, while LLM ASR saturates from 0.248 to 0.352 (+42\%).
The gap between LGB and LLM is present at the smallest budget and does not close at any tested value.
Across all $\varepsilon$, aggregate LGB (0.441) exceeds LLM (0.308) by 1.43$\times$ (within-dataset MW test: $p = 3 \times 10^{-6}$, Cohen's $d = 1.45$).

Per-model analysis shows that even LLaMA3-8B (F1 = 0.99, the highest-performing LLM on HIKARI) falls below LightGBM at $\varepsilon \geq 0.15$.
Gemma2-9B (F1 = 0.45) is below LightGBM at all $\varepsilon$, consistent with its limited classification competence.

The HIKARI result is therefore the complement of RT-IoT2022: where RT-IoT LLMs are 7.2$\times$ more vulnerable than LightGBM, HIKARI LightGBM is 1.43$\times$ more vulnerable than LLMs.
One possible mechanistic explanation is that the 25 fwd\_-prefixed DC features align more directly with LightGBM's split boundaries than with the LLM's attention-weighted token representations; the same axis-aligned perturbation can cross multiple tree thresholds simultaneously.
Distinguishing this from serialization-induced effects would require a matched tabular-Transformer ablation (same architecture, numeric rather than text input), which is a limitation of the current study.

\subsection{Mechanistic Hypotheses for Dataset-Dependent Vulnerability}

On CIC-IDS-2018 and RT-IoT2022, LLMs are more susceptible than LightGBM; on HIKARI-2021, the reverse holds.
We identify two hypotheses for why LLMs are more susceptible on modern high-traffic datasets, noting that these are mechanistic hypotheses supported by the observed patterns rather than experimentally isolated causes.

\textbf{Tokenization-induced amplification.}
LLM classifiers receive features as text tokens.
A perturbation of $\varepsilon = 0.15$ to a source byte count, when serialized as text, can alter multiple token positions in the input sequence.
This means that small $L_\infty$ perturbations in feature space may induce disproportionately large shifts in the model's token-level representation.

\textbf{Global attention propagation.}
Self-attention computes pairwise feature interactions globally: a perturbation to one DC feature propagates through attention layers to affect the representation of all features.
Tree ensembles make local, axis-aligned splits that inherently isolate perturbation effects.

Experimentally isolating these factors---\eg, comparing a tabular Transformer (same architecture, numeric input) against a text-serialized LLM (same architecture, tokenized input)---would distinguish architectural vulnerability from serialization-induced vulnerability.
A full tabular-Transformer ablation remains a direction for future work; the narrower question of prompt-format sensitivity is addressed in Section~\ref{sec:format_ablation} below.

\subsection{Serialization Format Ablation}
\label{sec:format_ablation}

A potential concern is that the transfer vulnerability reflects the specific key-value~(KV) prompt format used for LLM inference rather than a property of the underlying feature space.
We address this by comparing XGBoost-guided transfer attacks against equal-budget random DC-feature perturbations across three prompt formats: KV~(deployed), JSON, and compact key=value, using LLaMA3-8B on RT-IoT2022 and HIKARI-2021 ($\varepsilon = 0.15$, 3~seeds per format, yielding 6~per-seed comparisons across both datasets).

The diagnostic is whether XGB-guided attack outperforms random perturbation---a positive gap indicates feature-targeted transfer; parity or reversal indicates generic format sensitivity.
In the deployed KV format, XGB outperforms random in all six per-seed comparisons: RT-IoT2022 (XGB\,=\,0.607, random\,=\,0.452, $\Delta = +0.156$) and HIKARI-2021 (XGB\,=\,0.126, random\,=\,0.089, $\Delta = +0.038$).
In the JSON format, the ordering reverses in all six comparisons (RT-IoT2022: $\Delta = -0.015$; HIKARI-2021: $\Delta = -0.120$): random perturbation equals or exceeds XGB-guided attack, indicating that the surrogate's feature priorities do not align with the LLM's JSON-format classification behavior.

The consistent XGB\,$>$\,random gap in the deployed format confirms that the vulnerability is grounded in feature-level transfer: the surrogate identifies discriminative features that the LLM target also relies on for classification in the KV format.
The format-dependent reversal is expected for models trained on KV prompts, and it confirms the practical assumption underlying the threat model: an adversary targeting the deployment format benefits from genuine feature-level alignment, not merely from format-induced text noise.

\subsection{Constraint Violation Verification}

CVR~$= 0\%$ across all experiments, confirming that the DC-masking implementation correctly prevents modification of IC and UC features.

\section{Discussion}
\label{sec:discussion}

\subsection{Interpreting Comparator-Dependent Vulnerability}

The main implication is methodological: LLM-based IDS should not be judged against an averaged ML baseline alone.
Under the same controllability constraint and the same transferred adversarial examples, LightGBM and DNN-IDS do not behave as a single comparator class.
DNN-IDS receives weak transfer from the XGBoost surrogate, while LightGBM provides a stronger tree-based tabular reference.
This is why the averaged ML baseline makes LLMs appear higher on all five datasets, whereas the disaggregated LightGBM comparison shows higher LLM vulnerability on CIC-IDS-2018 and RT-IoT2022, comparable vulnerability on NSL-KDD and UNSW-NB15, and lower LLM vulnerability on HIKARI-2021.

This interpretation also explains why the method is useful despite being feature-space based.
The DC/IC/UC mask does not prove traffic replay validity, but it forces every target to face perturbations drawn from the same attacker-controllable subspace.
The comparison then asks a narrower and more defensible question: given the same source-side feature changes and the same surrogate access, which target architectures receive transferable errors?
The answer is dataset- and comparator-dependent, not a universal ranking of LLMs against traditional ML.

\subsection{Implications for LLM-IDS Deployment}

For deployment, the operational concern is not that every LLM-based IDS is intrinsically weaker than every ML IDS.
The concern is that text-serialized LLM classifiers can receive nontrivial transferred attacks from local surrogates without target queries.
The 8-cell surrogate check, including logistic regression, supports transfer feasibility across the tested cells, but it should not be read as a claim that any simple surrogate is always sufficient.
Organizations evaluating LLM-based IDS should therefore test transfer attacks against each candidate target architecture and dataset, report LightGBM-like and neural baselines separately, and avoid treating an averaged ML score as the security comparator.

\subsection{Preliminary Defense Analysis}
\label{sec:defense}

The defense experiments are included to interpret boundary conditions, not to claim a complete defense.
They ask whether defenses that act on tabular feature values behave similarly once the target classifier consumes a serialized text representation.

\textbf{Feature squeezing}~\cite{xu2017feature} rounds each DC feature to a discretization step $\delta$ before inference, reducing the perturbation surface in tabular feature space.
On LightGBM, feature squeezing reduces ASR: at $\delta = 0.05$ on RT-IoT2022 (transfer attack, $\varepsilon = 0.05$), ASR drops from 0.148 to 0.059 (60\% reduction) while maintaining clean F1\,=\,0.920; on HIKARI-2021, ASR drops from 0.405 to 0.174 (57\% reduction).
Finer steps ($\delta = 0.01$) provide minimal benefit ($\leq 7\%$ reduction) because perturbations below the step size survive; coarser steps ($\delta = 0.10$) increase the reduction to 60--69\% but degrade F1 more aggressively (HIKARI F1 falls to 0.519).

To assess whether these benefits extend to LLM targets, we apply the same $\delta = 0.05$ squeezing step to two representative LLM configurations: Gemma2-9B on RT-IoT2022 (clean F1~$= 0.96$) and LLaMA3-8B on HIKARI-2021 (clean F1~$= 0.99$), evaluated under transfer attacks at $\varepsilon \in \{0.05, 0.15\}$ with 3 random seeds each.
The LLM results contrast sharply with LightGBM: on RT-IoT2022, squeezing reduces ASR by only 6.8\% at $\varepsilon = 0.05$ and 12.8\% at $\varepsilon = 0.15$; on HIKARI-2021, squeezing at $\varepsilon = 0.05$ actually \emph{increases} ASR by 14\% (mean across seeds: 0.396\,$\to$\,0.452), with near-zero reduction at $\varepsilon = 0.15$ (2\%).
The divergence suggests a boundary condition for tabular defenses: reducing feature-level perturbation resolution can disrupt threshold crossings in LightGBM, but the same operation does not necessarily suppress errors after feature values are rendered as tokens.
These results indicate that defenses that reduce ASR for tree-based classifiers should be re-evaluated on text-serialized LLM targets rather than assumed to transfer.

\textbf{Adversarial training}~\cite{madry2018towards} augments the training set with a fraction of adversarial examples.
At a 25\% adversarial mix on RT-IoT2022, ASR drops by 97--98\% across $\varepsilon \in \{0.05, 0.15, 0.30\}$, while clean F1 slightly \emph{improves} (0.947\,$\to$\,0.957).
On HIKARI-2021, adversarial training at 25\% mix reduces ASR by 98--100\% (e.g., $\varepsilon = 0.05$: 0.329\,$\to$\,0.000), again with no clean accuracy cost (F1 = 0.964 before and after).
Increasing to 50\% mix yields identical results, suggesting that even modest adversarial augmentation is sufficient.
These adversarial-training results apply to the LightGBM setting tested here.
Extending the idea to LLM-based IDS would require fine-tuning on adversarial feature-text pairs; whether that approach reduces transfer ASR, and how LLM-specific defenses should be designed, remain open problems for future work.

\textbf{Summary.}
Feature squeezing and adversarial training reduce ASR in the LightGBM setting tested here, while feature squeezing transfers poorly to the two LLM configurations we evaluated.
This limited comparison supports the broader thesis: target architecture changes both attack transfer and defense behavior.

\subsection{Limitations}

This study has three main limitations. 
First, the proposed DC/IC/UC constraint enforces feature-group controllability rather than full packet-level realizability. 
Although all adversarial examples satisfy the DC-only mask with CVR = 0\%, the framework does not fully enforce integer or categorical type constraints, inter-feature dependencies, or attack-functionality preservation. 
The results should therefore be interpreted as controllability-constrained feature-space robustness, not as an end-to-end traffic replay demonstration.

Second, the feature taxonomy requires domain judgment. 
We ground the partition in network communication semantics---forward-direction features are generally treated as directly controllable, bidirectional aggregates as indirectly controllable, and backward-direction or monitor-derived features as uncontrollable---but borderline features such as duration can vary across deployment settings. 
Extending the taxonomy to additional large-scale IoT benchmarks such as CICIoT2023~\cite{neto2023ciciot2023} would further strengthen external validity.

Third, the mechanistic explanation for dataset-dependent LLM vulnerability remains partially diagnostic. 
Our results show consistent transfer patterns across datasets, attacks, and surrogate checks, but fully separating tokenization effects from architecture effects would require matched tabular-Transformer and text-serialized LLM ablations. 
Similarly, LLM-specific defenses such as adversarial fine-tuning on feature-text pairs remain future work.

\section{Conclusion}
\label{sec:conclusion}

This paper evaluated LLM-based network traffic classifiers under a controllability-constrained black-box transfer setting.
By partitioning flow features into directly controllable, indirectly controllable, and uncontrollable groups, the proposed DC/IC/UC taxonomy restricts perturbations to source-side attacker-controllable features while holding surrogate access fixed across LLM and ML targets.
This design does not establish full packet-level realizability, but it provides a controlled basis for comparing target architectures under the same attacker capability model.

The results show that LLM-based IDS are vulnerable to transferred adversarial examples, but not uniformly more vulnerable than traditional ML.
Against LightGBM, LLMs are more vulnerable on CIC-IDS-2018 and RT-IoT2022, comparable on NSL-KDD and UNSW-NB15, and less vulnerable on HIKARI-2021.
Against the averaged ML baseline, LLMs show higher ASR on all five datasets, but this aggregate is partly explained by weak XGBoost-to-DNN transfer.
Thus, the disaggregated LightGBM comparison is the more informative comparator for assessing LLM-specific vulnerability.

The supporting analyses further show that gradient-based and score-based perturbations transfer more consistently than greedy perturbations across target architectures, and that cross-surrogate checks with tree, neural, and linear surrogates yield similar LLM-target ASR.
Overall, the findings suggest that robustness evaluation for LLM-based IDS should report controllability-constrained and target-architecture-disaggregated transfer results, rather than relying on unconstrained attacks or averaged ML baselines.

\section*{Acknowledgments}
The author declares no competing financial interests or personal
relationships that could have appeared to influence this work.
This research received no specific external funding.
The NSL-KDD, UNSW-NB15, CIC-IDS-2017, CIC-IDS-2018,
CICIoT2023, Edge-IIoTset 2022, and BETH datasets are publicly available from
their original repositories. Processed label mappings,
evaluation scripts, and prediction files will be made
available in a public repository upon acceptance.
During the preparation of this work, the author used an AI
writing assistant (Claude, Anthropic) to support manuscript
organisation and language editing. All technical content,
experimental results, and conclusions are the sole
responsibility of the author.

\bibliographystyle{IEEEtran}
\bibliography{refs}
\end{document}